THE EUROPEAN
PHYSICAL JOURNAL C

Regular Article - Theoretical Physics

# Non canonical polarizations of gravitational waves

Stefano Bondani[1,a], Sergio Luigi Cacciatori[1,2,b]

[1] DiSAT, Università degli Studi dell'Insubria, via Valleggio 11, 22100 Como, Italy
[2] INFN, via Celoria 16, 20133 Milan, Italy

**Abstract** We hereby propose an alternative and additional angle on the nature of gravitational waves (GWs), postulating the theoretical and experimental possibility that GWs carry a deformation of the time component of spacetime, other than the spatial one. By explicitly working outside of the transverse-traceless gauge, we propose how events with well-defined time duration, when hit by a GW, would consequently be expected to show a difference in their characteristic time, as measured from the rest frame of an outside observer, whose clock is to remain unaffected by the GW. This constitutes a theoretically viable way in the sense of detecting the passing of the wave itself and may prove relevant as a standalone method for GWs detection other than laser interferometers, or as well be implemented as a complementary but independent system of signal triggering, improving the statistical significance of existing methods. A simple but physically realistic scenario in which the appropriate conditions for the generation and detection of GWs with time dilation are met is presented, along with the conceptual design of an experimental detector.

## 1 Introduction

The detection of GWs with laser interferometers [1,2] has proven immensely successful for several years now [3–6]. The phase shift of laser beams measured by interferometers originates from a variation of the time flow induced by the GW. Because in any locally inertial reference frame the speed of light $c$ is constant, the phase shift is then translated in the contraction and expansion of the proper distance, or, in other words, the length of the interferometer's arm. For instance, in the transverse-traceless (TT) gauge, spatial coordinates are left invariant at first-order, while what changes is the laser's proper travel time, i.e. its phase as measured by an observer for whom the speed of light is equal to $c$.

[a] e-mail: sbondani@uninsubria.it (corresponding author)
[b] e-mail: sergio.cacciatori@uninsubria.it

The problem of associating coordinates' measurements to physical variables is in general not a trivial one, and especially arduous in the context of GW detection. Spacetime is, in a sense, *undivided*, and yet we usually expect to be able to isolate a particular observable, a spatial or time coordinate, as if it were an independent quantity.

When hit by a GW, in the language of special relativity, an event in a 3+1 $D$ spacetime is subject to a strain, usually parameterized in linearized general relativity formalism by $h_{\mu\nu}$, which, assuming the idealized situation of adopting a weak field approximation and of being in an inertial frame, is assumed as a perturbation of a flat Minkowski metric $\eta_{\mu\nu}$ as

$$g_{\mu\nu}(\mathbf{x}) = \eta_{\mu\nu} + h_{\mu\nu}(\mathbf{x}). \qquad (1)$$

Keeping in the weak field approximation, the strain $h_{\mu\nu}(\mathbf{x})$ propagates in vacuum ($T_{\mu\nu} = 0$) as a wave,

$$\Box h_{\mu\nu}(\mathbf{x}) = 0 \qquad (2)$$

at speed $c = 1$. In order to make explicit the existence of a GW in the most straightforward way, it is fairly common practice to move onto the TT gauge

$$h^{0\mu} = 0, \quad h^i{}_i = 0, \quad \partial^j h_{ij} = 0, \qquad (3)$$

which identifies the metric $h_{ij}^{T,T}$. The "transverse" condition, in particular, means that the maximal strain effects reside in the planes orthogonal to the direction of propagation of the GW; in other words, in this specific gauge, the GW oscillations are orthogonal to the 4-vector $U^\mu$ of an observer moving towards the propagation direction of the GW, i.e.

$$A_{\mu\nu} U^\nu = 0 \qquad (4)$$

(with $A_{\mu\nu}$ the generic GW amplitude); this alone does not say anything about possible lesser effects at different angles with respect to the propagation direction. Moreover, one should also try to be more explicit about the meaning assigned to the TT gauge itself: as was indeed pointed out in [7,8] the notion of transverse-traceless might refer either to a decomposition of the metric perturbation $h_{ij}$ which is local in momentum space and non-local in physical space or to that part of $h_{ij}$ obtained with a projection operator $P_i{}^j$ local in physical space. The two notions are only interchangeable under specific conditions and are in general not equivalent. Once the appropriate notion of TT gauge is adopted, the two independent constant amplitudes, $h_+$ and $h_\times$, can be derived, representing the possible polarizations of the GW, which are at an angle of $\pi/4$ to each other and – it's worth pointing out – emerge from and after choosing the gauge. On a side note, it was shown [9,10] that the canonical double polarization of GWs is not the sole possible solution as, for instance, spin-1 GWs might be obtained as exact solutions of Einstein's equations (with cosmic strings or gamma ray bursts as sources). Although tentative, this result suggests a possibility in the sense of non-canonical GW polarizations and observables when working outside the constraints of the TT gauge.

More broadly, it's worth noting some recent attempts carried out by the LIGO-Virgo Collaboration (LVC) [11], in the direction of a search of signatures beyond GR in their GW signals, where without relying on any specific theory of gravity, a search for possible GWs with up to six different polarizations was performed assuming a generic metric theory. Although inconclusive, this work showed that the possibility of finding new solutions outside the canonical framework of the GR+TT gauge is not completely out of the equation, and could still be worth pursuing.

Let's start by considering that during the passage of a GW what gets displaced are not just test masses over a spatial distance but the entire spacetime interval

$$ds^2 = g_{\mu\nu}(\mathbf{x})dx^\mu dx^\nu \quad (5)$$

between events in the perturbed spacetime. Indeed, as GWs traverse spacetime they are locally modifying the gravitational field through perturbations of the Riemann tensor $R_{\mu\alpha\nu\beta}$: not only distances are affected from the point of view of an external observer but times are also expected to do the same. Although the existence of a GW is usually derived from the $\partial_\mu \bar{h}^\mu{}_\nu = 0$ condition, an equally valid criterion is to work with the Riemann tensor itself. In fact, during the passing of a GW, Eq. (5) can be recast in a more appropriate form – using gaussian coordinates – as

$$ds^2 = \left(\eta_{\mu\nu} + \frac{1}{3}R_{\mu\alpha\nu\beta}x^\alpha x^\beta\right)dx^\mu dx^\nu. \quad (6)$$

## 2 The asynchronous traceless gauge

More explicitly, it might prove useful to relax the usual TT gauge approach and formulate a hypothesis on what could be a possible alternative physical observable resulting from this choice; it's conceivable that this could lead to the emergence of one or more different polarizations of the GW. In this sense, our claim is that both approaches are correct, but explanations of the different emerging observables could vary based on the gauge choice. It is equally important to point out that whatever the perturbed metric will be, it will be essential to express it in terms of the Riemann invariant (at a linearized level) tensor, rather than the Christoffel symbols.

For example, a minimal modification of the gauge choice could be the following one. After fixing the Landau–Lorenz gauge, we can use one of the four residual degrees of freedom to impose the vanishing of the trace. Then, there remain 3 further degrees of freedom commonly employed to impose the vanishing of the $h_{0j}$ terms. Before imposing such constraints, for a wave $h_{\mu\nu}(z - ct)$ moving along the direction $z$, we have

$$h'_{00} + h'_{30} = 0, \quad (7)$$
$$h'_{03} + h'_{33} = 0, \quad (8)$$
$$h'_{01} + h'_{31} = 0, \quad (9)$$
$$h'_{02} + h'_{32} = 0, \quad (10)$$

where a prime indicates derivative w.r.t. the argument, and 3 is associated to $z$. The first two equations, together with the traceless condition, show that we must have $h_{00} = h_{33}$ and $h_{11} = -h_{22}$. From the last two equations, we see that if we use two of the residual freedoms to fix $h_{01} = h_{02} = 0$, then also $h_{31} = h_{32} = 0$. At this point, we are left with the waveform

$$h_{\mu\nu} = \begin{pmatrix} h_o(w) & 0 & 0 & -h_o(w) \\ 0 & h_+(w) & h_\times(w) & 0 \\ 0 & h_\times(w) & -h_+(w) & 0 \\ -h_o(w) & 0 & 0 & h_o(w) \end{pmatrix}, \quad (11)$$

where $w = z - ct$, and still one gauge freedom remaining. Usually, this last gauge freedom is used to put $h_o = 0$, but we can instead use it to impose $ah_\times = bh_+$ for fixed $a, b \in \{0, 1, -1\}$ not both vanishing. For example, we could impose the vanishing of the + mode, thus remaining with the × and the $o$ modes. In particular, the latter corresponds to the impossibility of keeping clocks along the $z$-axis synchronized during the passage of the GW.

It's clear that in this gauge, which is manifestly a-synchronous, the *time-time* component of the amplitude, rather than being reabsorbed inside the $h_{+,\times}$ components of the strain, as it happens with the standard TT gauge, survives and can therefore be treated as an observable. We can call this gauge choice the A-synchronous Traceless (AT) gauge (having only one transverse mode and a "desynchronization" mode). The intuition behind the two waveforms emerging from the different gauges being different representations of the same wave comes from the equivalence principle.

Moreover, the class of diffeomorfisms $\xi^\mu(t, x, y, z)$ such that

$$\partial_\mu \xi_\nu + \partial_\nu \xi_\mu = -\delta g_{\mu\nu} \tag{12}$$

cancels the $+$ or $\times$ mode, automatically implies that a mixing of the coordinates in the dependency of $h_o$, instead of the simple $w$, must take place. Therefore, if such dependency is on terms containing the $x$ and $y$ coordinates, the desynchronization happens not just along the $z$ axis, but on the $xy$ plane as well, with different amplitudes as function of $x$ and $y$. This implies that the general desynchronization front, mapping the degree of clock de-phasing at each point, is a curved surface, rather than a plane, characterized by its own iso-desynchronization gradient lines, much like an equipotential diagram. By preventing spatial deformations along the $x$ axis, clocks will be desynchronized on that same axis. In other words, the TT-gauge spatial deformations seen by an interferometer, where clocks are fixedly synchronized, are being reabsorbed in the desynchronization of clocks in the AT gauge, where spatial contractions in the corresponding directions are impeded.

## 3 Time dilation as a technique for the detection of gravitational waves

The order of magnitude of $h_{\mu\nu}$ for astrophysical sources commonly falling in the LVC[1] band (stellar-mass black hole or neutron star mergers from some $10^{2 \div 3}$ Mpc luminosity distance) is generically assumed as a target for the sensitivity of laser interferometers, in that

$$\frac{\Delta l}{l} \simeq h_A \tag{13}$$

with $A$ the $+$ and $\times$ polarizations (e.g., in the TT gauge), and $l$ referring to one of the spatial coordinates (e.g. the length of one of the two arms of the interferometer).

For a typical LVC-like event – say, a binary stellar black hole merger in the kHz – with a total burst duration (measured from Earth) of a few $10^2$ ms, this would be the time interval over which the spacetime dilation is taking place and during which any measurement of distances and/or times should show some difference from the unperturbed conditions. Distance changes are what is currently observed with laser interferometers, while time changes are what is being proposed here.

Of course, a more persistent GW signal at lower frequencies, e.g. in the LISA band [12–14], or even in the $\mu$Ares [15] $\mu$Hz frequency band, would yield similar effects over longer timescales.

During the first three LVC campaigns, the strain $h_A$ of confirmed GW events have been observed to show values usually between the order of magnitude of the instrumental sensitivity $S_n \simeq 10^{-23}$ and in the excess of $1.0 \times 10^{-18}$ [4]; the design sensitivities of advanced LIGO [1,18] and advanced Virgo [19] currently exceed $10^{-23}$ at frequencies $\approx 10^2$ Hz.

Our proposal is that if $h_{\mu\nu}$ is assumed to be acting over an unperturbed spacetime, one may then expect it to have a similar, if not the same, effect over a time interval, when measured by an external observer, i.e.

$$h_A \simeq \frac{\Delta t}{t} \simeq \frac{\Delta l}{l} \simeq 10^{-23 \div -18} \tag{14}$$

where $t$ is the coordinate time measured by an observer at infinity. The crucial point is that the proper time $\tau$ measured by a physical observer in the same rest frame of an event crossed by the trajectory of the GW will of course notice no alteration in its own rate. We point out that this is no different from two observers with identical clocks, one of them close to the event horizon of a black hole, the other at infinity, at rest in a flat Minkowski: while the two observers in their reference frames will observe their clock to progress at a rate of 1 s per second, the observer at infinity would, in principle, be able to observe the other clock in the proximity of the black hole advancing at a slower rate. Similarly, the observer near the black hole would observe the clock at infinity proceeding at a faster rate. Naturally, the assumption behind Eq. (14) is yet to be rigorously derived and proved. If this is indeed true, then what happens is that over timescales much shorter than the duration of the event itself ($\simeq 0.1$ s, for an event like GW150914), the proper time contracts and expands in phase with the frequency of the GW with an amplitude given by $h_A$, but only an external observer unaffected by the GW would be able to detect such contractions. It is only intuitive to expect these contractions and expansions to cancel out over each period of the GW sinusoid.[2]

To restate the idea from the reference frame comoving with the GW, inside a stronger gravitational field, given by a positive GW strain, time will be slowed down, and likewise, it will be during both the ascending phase of single crests of the GW sinusoid – until the local maximum is reached where time will flow at its slowest rate – and the descending phase, where it starts accelerating back, until crossing the zero of the curve, where no time dilation happens. Similarly, when the strain is negative, during the descending phase, time flow

---

[1] If and where applicable, LVC in the text is intended equivalent to LVK, i.e. the LIGO-Virgo-KAGRA Collaboration [18,19].

[2] The amplitude of the signal is not exactly constant though, more so during the final merger phase of a binary source; therefore one should not have exact cancellation of the signal between periods, but at least a residual effect in the measured time, analogous to what happens with the Christodoulou memory [20–24].

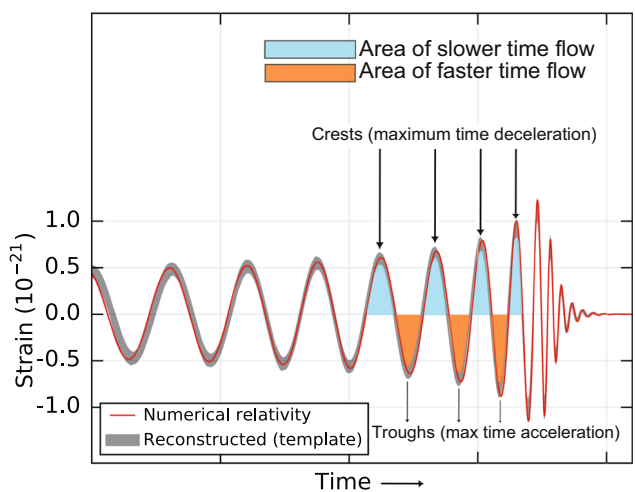

**Fig. 1** Areas in blue (orange) show the time intervals over which the strain is positive (negative) and over which time measurements should show a deceleration (acceleration), meaning in other words that a millisecond should take a millisecond plus (minus) one part in the order of magnitude of $h_A$, say $10^{-21}$, when measured by an observer at infinity. (Figure is an adaptation from [3])

will accelerate and at the local minimum, e.g. at the trough, time will flow at the fastest rate; during the ascending phase it will slow down again until the next zero; the process will then repeat for the subsequent period. Figure 1 is adapted from [3] and makes an attempt at picturing the idea: where (or better, *when*) the strain is positive (colored in blue), the metric tensor is positively perturbed, meaning a stronger gravitational field, and a slower flowing of time when measured from an external observer. The vice-versa holds for negative strains (colored in orange).

### 3.1 Conceptual difference with pulsar timing arrays and Doppler shift tracking

It would be tempting at this point to see a strong similarity between this concept and what is generally referred to as pulsar timing array (PTA) [25–30]. PTAs observe the incoming signal from distant pulsars, which are treated as highly regular clocks, and measure the time interval between the incoming pulses. Under unperturbed conditions the $\Delta t$ between the pulses is constant; deviations from this regularity are interpreted as the passing of a GW between the source and the observer on Earth. The GW is therefore acting on the geodesic between them, resulting in a different travel time for the photons emitted by the pulsar, hence a shift in their time of arrival.

By contrast, what we are assuming here, and which will be detailed in Sect. 4.2, is a GW acting on a source itself of electromagnetic signal, and somehow not on the subsequent null geodesic covered by any EM signal coming out of it: it's the time dilation happening at the source to cause any subsequent measure of $\Delta t$ at the observer rest frame, rather than an intervening GW on the path of a signal which originated regularly in unperturbed conditions at the source. In other words, it is a differential, rather than integral, discrepancy.

A further possible misconception may arise with the notion of spacecraft Doppler tracking [31], where GWs act over the on-Earth observed Doppler shift of sinusoidal Electromagnetic signals to and from a spacecraft. Again GWs are acting on the photons' propagation, varying the metric along the signal path. The similarity with our proposal resides in the requirement for clocks to be as precise as the GW strain, i.e. with precision better than 1 part in $10^{18}$ for most sources. Yet again the difference is centered on the influence of the GW over the spatial, rather than the temporal component of the metric.

In a sense, a higher conceptual similarity can be found with the gravitational Aharonov–Bohm effect [32], namely the phase shifts induced by the gravitational influence of a close-by mass over proper time differences between freely falling, nonlocal trajectories in atom interferometers [35,36]. The difference, in this case, is in the nature of the source, a mass, rather than a GW.

## 4 Towards an experimental observatory

Conceiving an experimental device capable of measuring time dilations in a way that can be useful for GWs detection is by itself quite a challenging task. The main difficulty is to have the GW influencing the timing of a physical process but not the clock measuring it. Furthermore, for any such detector to be conceptually different from an interferometer, one must make sure that the influence over any time measurement exerted by the GW doesn't stem from the spatial component being perturbed, altering the travel time of the signal.

Let us assume to possess a sample where some particular physical process is known to happen with characteristic time intervals $\tau^*$, e.g. a perfect clock emitting particles at a constant rate; for instance, Wilczek's time crystals [37–41] could – at least ideally – serve this purpose. In theory, if such features are then continuously observed and therefore timed, being their $\tau^*$ known a priori, one should observe a different behavior, namely of the order of $h_A$, in terms of elapsed times when a GW is passing by the samples and altering their characteristic time when measured by an observer at infinity. The main difficulty is once again to have the external observer somehow unaffected by the passage of the GW itself, at least not during the time measurement of the process influenced by the GW, or in other words to be non-local in physical space with respect to the reference frame of the samples; if this were not so, the external observer's clock would be in turn affected by the GW, and although present, the phenomenon could not be detected.

A sufficiently precise clock to serve as detector is therefore the first prerequisite towards this program, and is the subject of the next subsection, where we will refer indistinctly to precision or resolution of the clock.

### 4.1 Timing resolutions

When dealing with a LVC-like GW event in the kHz occurring altogether over a few $10^{-1}$ s (as is roughly the case for GW150914) and $h_A \simeq 10^{-23 \div -18}$ it follows that the term $\Delta t$ in Eq. (14) has to fall in the range of the $10^{-24 \div -19}$ s; if dealing with a single period between crests of the wave (or more accurately between zeros of the amplitude), which can be assumed to happen over a few $\simeq 10^{-2}$ s [3], one finds a requirement for $\Delta t \simeq 10^{-25 \div -20}$ s. This is the precision required from the measuring clock for an event of this kind. Table 1 lists some of the shortest characteristic timescales for processes known or predicted by particle and nuclear physics. The fact that such short timescales are predicted or inferred by fundamental physics does not straightforwardly nor automatically imply if and how they could be put to use in this context.

More attainable time measuring systems include processes of interactions of quick laser bursts with matter, with current technology reaching periodic times between pulses in the region of $10^{-18}$ s [42]. Similar timing resolutions in the attosecond have recently been claimed in high-harmonic spectroscopy with Gouy phase interferometers [43,44]. Recent claims of timing precision measurements with uncertainty down to $9.7 \times 10^{-19}$ obtained with optical atomic clocks based on quantum-logic spectroscopy [45], or even down to $2.5 \times 10^{-19}$ obtained with imaging spectroscopy techniques [46], place the aforementioned required range inside current technological capabilities. A timing resolution around this order of magnitude would be more than adequate when trying to detect GWs at lower frequencies than those in the LVC band. For a GW event falling in the LISA band, with frequency, say, of $\simeq 10^{-2}$ Hz, the actual required $\Delta t$ in Eq. (14) would be $\simeq 10^{-19 \div -16}$ s.

**Table 1** List of some of the shortest known or predicted characteristic timescales of physical processes

| Duration [s] | Process |
| --- | --- |
| $0.3 \times 10^{-24}$ | Mean lifetime of the **W**± and **Z** bosons |
| $1 \times 10^{-24}$ | Top quark decay |
| $1 \times 10^{-24}$ | Gluon emission from a Top quark |
| $9.1 \times 10^{-23}$ | Half-life of $^4$Li |
| $7 \times 10^{-21}$ | Neutron half-life in $^9$He nuclear halo |

It is worth mentioning that in 2021 the quantum phase of a collectively excited nuclear state was tuned via transient magnons with a precision of $1 \times 10^{-21}$ s, monitored interferometrically via quantum beats between different hyperfine-split levels [47]; moreover, in 2022 the gravitational redshift within a single millimeter scale sample of ultracold strontium was measured with a fractional frequency measurement uncertainty of $7.6 \times 10^{-21}$ [48]. It is clear that a timing resolution in the zeptosecond or even higher would enable studying spatial gradients of the time clock, thus opening the door for a detailed waveform reconstruction and GW astrophysics of the source.

All this being said, the actual scheming of an ideal device capable of arranging the samples and the observer so that they could firstly be sensible to a GW in terms of time dilation, and secondly be detectable by the external observer is not a trivial task and will be tentatively addressed in the next subsection.

### 4.2 Concept design

Let us assume to have at an initial time $t = 0$ (Fig. 2, left panel, not to scale) an incoming GW with wavelength $\lambda$, an ideal localized source (labeled 'E') of size $d$, emitting particles – say, electrons or photons – at a fixed rate, towards a detector (labeled 'D') placed at fixed distance $R$, consisting of a clock measuring their $\Delta t$ of arrival. Let the system be in free fall, i.e. fully inertial, in an empty flat Minkowski background spacetime, where the curvature radius of the background can be written as:

$$\mathcal{R}_\mathrm{B} \equiv \left| \frac{1}{\sqrt{R^\mathrm{B}_{\mu\alpha\nu\beta}}} \right|. \qquad (15)$$

Let's further assume by now to consider a single crest (or trough) of the GW, rather than the entire waveform of several periods, and to evaluate its effect on the system as a standalone GW wavelet, or rather, a solitonic GW [49,50]. The first requirement for such a system would be for it to observe the condition:

$$R \gg \lambda \gg d. \qquad (16)$$

Now the idea would be to have at $t = 1$ (Fig. 2, middle panel) the GW sweeping past the entire system, including the detecting clock, in a finite and short enough time, and the signal emitter to possess its own internal clock, related to a highly predictable physical process, determining the emission rate of the particles/photons. The crucial requirement is for the signal to be *generated* while under the influence of the GW, but to be *expelled* at a retarded time with respect to the transit of the entire GW, so as to exclude any contraction effect over the path length between the emitter and the detector.

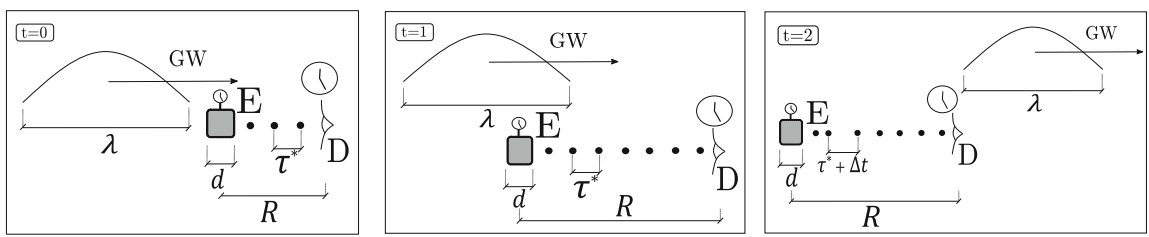

**Fig. 2** Schematic representation (not to scale) of an ideal detector, labeled "D", and an emitting source of particles, labeled "E", at times $t = 0, 1, 2$, i.e. before, during, and after the transit of a single GW crest, respectively. Note that the geometry of the system $(R, d, \lambda)$ does not change over time, and the sizes or shapes in these drawings are purely qualitative

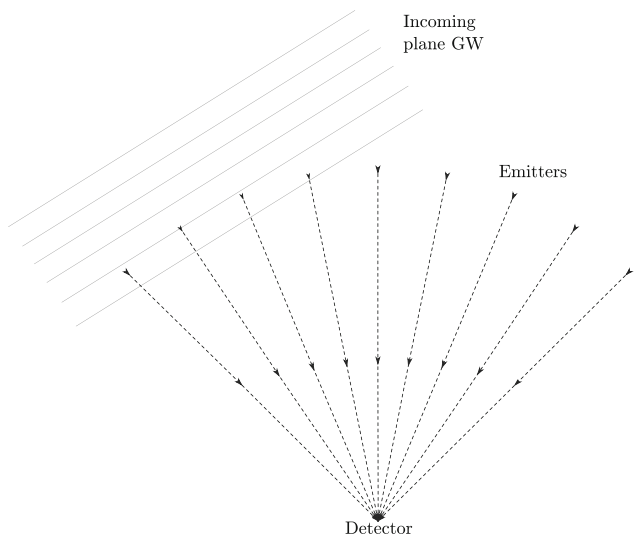

**Fig. 3** Overhead view of a basic distribution of emitters with a common ideal detector hit by an incoming GW. The time dilation will be observable only after the GW has left the system

This delay is required to be long enough so that when the particles are finally expelled ($t = 2$, right panel of Fig. 2), they do so in a flat Minkowski background once again, since the GW has already left the system, but crucially encoding a difference in the $\Delta t$ between their emission times, previously registered inside the device while under the influence of the GW. Equally essential is the requirement that a physical signal is actually produced, although still *inside* the device itself, only to later be expelled outside of it, with the strong caveat that the interdistance between each individual signal is to be accurately preserved, by an appropriately devised spiraling or mirroring internal system. Because of the delay between generation and expulsion, entailing by design absence of gravitational radiation during the travel of the signal from emitter to detector, the particles can have in principle any speed, and massive particles, as well as photons, can be employed.

If the signal is continuously launched outside of the device, it will not encode straight away the time dilation carried by the GW on the physical source inside of it, but such signal will be affected nonetheless by the GW crossing its path, much like inside an interferometer or with PTAs. The information on time dilation alone will come later on when the GW has already left the system. The amount of delay between production and expulsion of the signal is the designer's choice and hence determines the GW frequency range available for detection: the longer the target GW wavelength, the greater the required delay will need to be, and viceversa. As long as the technological limitations on clocks' precision are within practical employability, this system is scalable to cover a wide range of GW frequencies. In principle there is no need to increase the number of observatories to cover different GW frequency ranges: all that is needed is to univocally flag the signals emitted at different delay times for the detecting clock to distinguish between them.

### 4.3 More complex configurations

A natural next step toward a feasible experimental observatory would be to have several of the aforementioned systems, possibly in some sort of distribution equidistant from and directed towards the same detector, let's say an atomic clock of arbitrary precision capable of keeping memory of the time of arrival from each distinct signal-emitting device. Recent results in elementary quantum networks of entangled optical clocks [51] might suggest a possibility in this sense. A simple two-dimensional distribution of particle emitters, all oriented towards a common detector, is depicted in Fig. 3. An incoming GW wavelet hitting the system of emitters at an angle, would, at first, influence the timing of particle emission of some, and only some, of them; only at a later time would all the emitters be influenced by the GW, entailing a rough estimate of the direction of propagation of the GW.

A further development of our concept design would be to have a distribution of samples in a 3D spatial volume, emitting signals towards a central detector, timing the difference of arrival time from the different directions. Having them all at the same distance from the center would result in a spherical configuration like the one in Fig. 4.

This would produce a gradient of proper times, varying in phase with the transit of the GW, which could then be used to better reconstruct the direction of propagation of the GW, and therefore infer the sky localization of the source. At this stage the entire concept is almost completely scale-free, and the only proper requirement is for the size of the emitting device to be as point-like as possible, when compared with the GW wavelength. The delay between production and expulsion, makes the $R \gg \lambda$ condition rather less stringent than the first. The geometry of the system, in other words, is largely independent from the observed GW frequency.

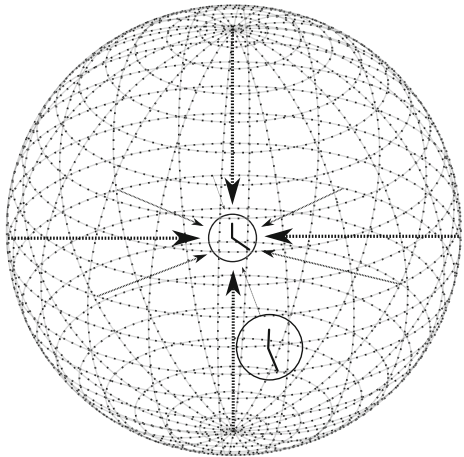

**Fig. 4** Schematic view of a complete three dimensional spherical detector. The samples, ideally free falling in flat Minkowski space (neglecting for now any mutual gravitational interaction), are disposed on the "surface" of the sphere: they produce end expel signals at a constant and known rate, towards the central receiving clock, which registers their ToA. When the received signal is not isotropic, and the anisotropy has a clear non static pattern moving across the entire system, a (single crest) GW might be responsible. The delay between production and expulsion of the signals from the samples guarantees that the observable is the GW-induced time dilation, rather than the path length contraction

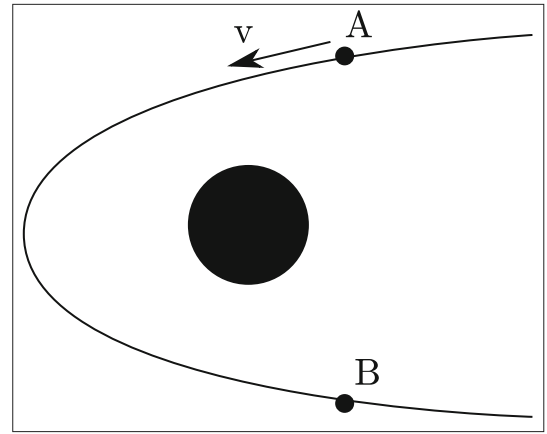

**Fig. 5** Extreme mass ratio collision between a test particle and a massive body. The collision is defined to begin and end on points A and B on the particle's trajectory

At the *gedankenexperiment*-level, an even more ideal device capable of conserving and preserving an arbitrary amount of information, could in principle be able to produce enough EM signal to cover the entire duration of a GW event, i.e. to map a greater portion, let's say the final few periods, of the GW sinusoid. To keep things reasonably simple, in the remainder of this work we will still limit to the case of single crest/trough GWs.

### 4.4 Detectability of a galactic signal

In order to construct a physically realistic scenario in which a single GW crest (or trough) can be generated, we will assume the simple astrophysical case of an elastic collision, depicted in Fig. 5, between a test particle and a massive body, or an extreme mass ratio collision (EMRC), as our GW source. In particular, let's assume the massive body to have mass $m_1$ equal to that of Sagittarius A$^\star$ (Sgr A$^\star$, $m_{\text{Sgr A}^\star} = 4.3 \times 10^6 \, M_\odot$ [52]), the supermassive black hole at the Galactic

Center, and the particle to be a compact object, for instance, a primordial black hole (PBH) with $m_2 = 1 \, M_\odot$ [53–59], coming from past infinity, and directed to future infinity after the collision, which we define to be restricted between point A and point B in Fig. 5. Because of the high mass ratio, we can safely consider both the fixed scattering center and the center of mass of the two-body collision to be located inside the supermassive black hole.

The initial and final 4-momenta of the PBH are $P^\mu$ and $P'^\mu$, respectively. Following [60,61] on the formalism of GW generation from two-body collisions, and from now on indicating with $\omega$ the GW frequency, the 4D Fourier transform of the corresponding energy-momentum tensor $T^{\mu\nu}(\mathbf{x}, t)$ in the non-relativistic limit is:

$$\tilde{T}_{ij}(\omega) \simeq -\frac{ic}{m\omega}(P_i P_j - P'_i P'_j). \quad (17)$$

The non-relativistic limit is valid for relative velocity between the two bodies $v \lesssim 0.3 \, c$, a condition which is respected when assuming an impact parameter as small as $b = 2.5 \times 10^{-6}$ pc $= 6 \, R_S$. We impose as a safe initial condition $v = 0.15 \, c$, resulting in a collision total duration of approximately $t_c \approx b/v$. The collision can be approximated as instantaneous so that the spectrum needs to be cut at frequency

$$\omega_{\text{max}} \sim \frac{2\pi}{t_c} \sim \frac{2\pi v}{b} = 3.7 \times 10^{-3} \, \text{Hz}. \quad (18)$$

If the collision is happening in the $(x, y)$ plane, most of the gravitational radiation will be emitted along the $z$ direction, and will be described by an angular distribution with polar coordinates $(\theta, \phi)$. If the test particle is scattered by the massive body by an angle $\vartheta_s$, the non-relativistic energy spectrum integrated over the solid angle $d\Omega = d\cos\theta d\phi$ is given by:

$$\int \frac{dE}{d\omega d\Omega} d\Omega = \frac{dE}{d\omega} = \frac{8G}{5\pi c^5} \mu^2 v^4 \sin^2 \vartheta_s, \quad (19)$$

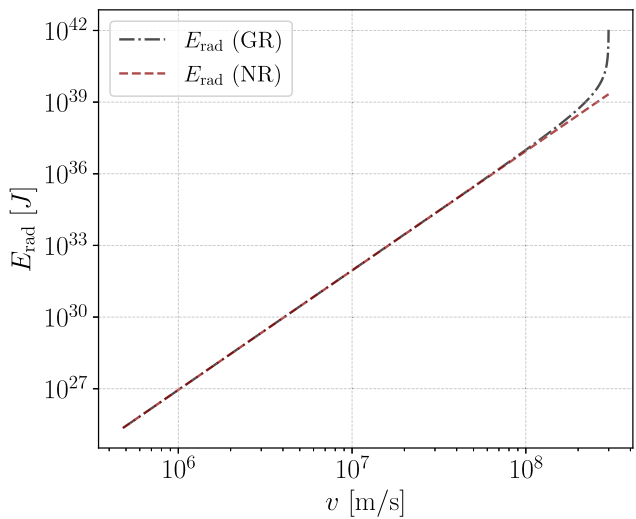

**Fig. 6** Total radiated energy after the collision. The discrepancy between the full general relativistic and non-relativistic limit becomes evident only at $v > 0.3\, c$

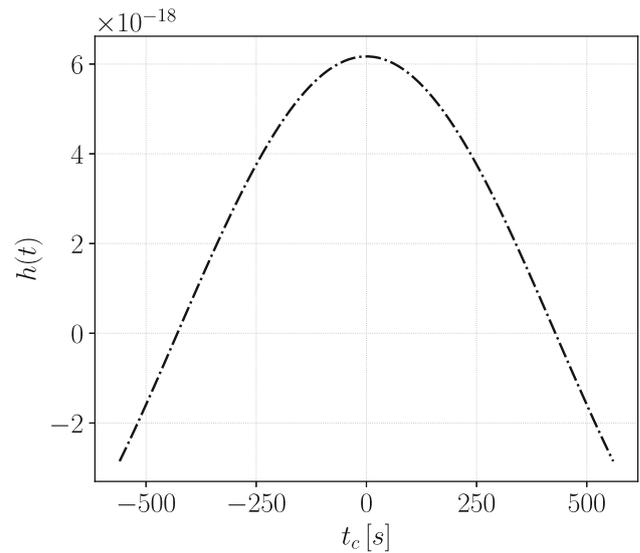

**Fig. 7** Waveform resulting from a collision with our chosen initial conditions. The approximation holds true only when the argument of the cosine does not exceed a certain threshold, which implies conditions on both $v$ and $t_c$

where $\mu = (m_1 m_2)/(m_1 + m_2)$ is the reduced mass. Integrating Eq. (19) up to $\omega_{\text{max}}$ gives the total radiated energy in GWs after the collision:

$$E_{\text{rad}} \sim \frac{16 G \mu^2}{5 b} \left(\frac{v}{c}\right)^5 \sin^2 \vartheta_s. \tag{20}$$

In the general relativistic case, Eq. (17) is replaced by a full expression for $\tilde{T}^{\mu\nu}(\mathbf{k}, \omega)$, with the 4-momentum containing a Lorentz factor, and the expression for the angular distribution of the energy spectrum takes a more complicated form with additional factors expressing the bending of the radiation pattern in the direction of motion. Figure 6 shows the total upper limit in total radiated energy as a function of relative velocity for the GR and non-relativistic cases, by assuming the scattering angle to be $\vartheta_s = \pi/2$; the result is expressed in Joules. The discrepancy between the two curves starts being significant only for values of $v > 0.3\, c$, which justifies the choice for adopting the non-relativistic limit.

Expressing the (average) luminosity $\mathcal{L}$ as simply $E_{\text{rad}}/t_c$, the relation between luminosity and energy flux is given by:

$$\mathcal{F} = \frac{\mathcal{L}}{4\pi D_L^2}, \tag{21}$$

where for the luminosity distance $D_L$ from Sgr A$^\star$ we adopted $8.26 \times 10^3$ pc [52].

To obtain a first-order raw estimate of the expected GW strain, we plug this into the expression for the energy flux from [60], which strictly would only apply after averaging over several periods, which of course don't take place during a one-time collision, but can nevertheless be instructive about the expected magnitude of the result:

$$\mathcal{F} = \frac{1}{32\pi} \frac{c^3}{G} h_A^2 \omega_{\text{GW}}^2. \tag{22}$$

Finally, substituting $\omega_{\text{max}}$ for $\omega_{\text{GW}}$, gives an estimate for the generic polarization strain:

$$h_A \simeq 6 \times 10^{-18}. \tag{23}$$

Similarly the expression for the time derivative of $h$ is given by [60]:

$$\mathcal{F} = \frac{1}{32\pi} |\dot{h}|^2 \frac{c^3}{G}, \tag{24}$$

giving an estimate for $\dot{h} \simeq 2 \times 10^{-20}$. Keep in mind that these results are obtained for a scattering angle $\vartheta_s = \pi/2$ that maximizes the total radiated energy in Eq. (20). Different values of $\vartheta_s$ will ultimately entail a somewhat smaller value of the GW strain, but the order of magnitude of the results is largely preserved for a wide range of scattering angles. With the aforementioned initial conditions and time duration for the collision $t_c \approx 1$ ks, one can try and give a zeroth-order estimate for the waveform, once again borrowing improperly from the quadrupole formalism, which simply reads:

$$h(t) = h_A \cos(\omega_{\text{GW}} t), \tag{25}$$

where substituting again $\omega_{\text{max}}$ for $\omega_{\text{GW}}$ and letting $t$ vary from 0 to $t_c$, we find the resulting truncated waveform of Fig. 7.

Since we are interested in the generation of a single GW crest (or trough), the approximation holds true only as long as the argument of the cosine in Eq. (25) doesn't exceed a certain threshold, or in other words, a full sinusoid is not able to form. Because we defined $\omega_{\text{GW}}$ in terms of the relative velocity $v$, this results in a constraint on the degeneracy of the two parameters $v$ and $t_c$ appearing inside the cosine. The shaded area in Fig. 8 shows the points in such parameter space where this condition is observed, i.e. the resulting sinusoid doesn't exceed a half period; the edge, in particular, indicates the points where the waveform – when centered on $t = 0$ – is truncated exactly at the change of sign of its second derivative, which maximizes the integrated GW strain, or in our language, the time dilation.

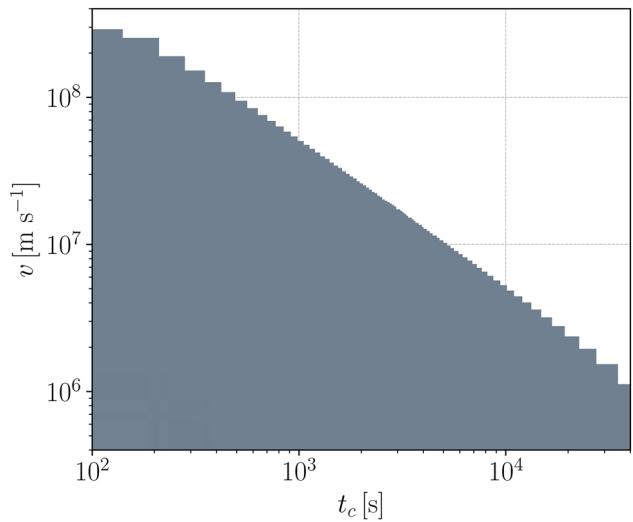

**Fig. 8** Parameter space showing the possible values of $v$ and $t_c$ for which only a single crest or trough of the GW waveform can develop

Of course the closer the points are to the edge of the shaded area, the more developed the GW crest (or trough, with a simple phase shift of $\pm\pi$) will be; the viceversa holds true. Conversely, points just outside of it are still permitted, but will result in a sinusoid with components of the opposite sign, inducing an increasing degree of cancellation of the time dilation produced by the GW; total cancellation will happen when a full period of the sinusoid is formed (see footnote 2 on page 2).

## 5 Conclusions

Detection of GWs with laser interferometers is nowadays an affirmed technique. Nevertheless, a second path towards reaching a similar result taking advantage of unexplored aspects of the same underlying physics may be permitted. The spacetime perturbation carried by the GW acts in principle both over spatial as well as time intervals, the only invariant ultimately being simply the first order Riemann tensor, and the observables being themselves a choice emerging as a result of the adopted gauge. By means of defining an Asynchronous-Traceless gauge, in lieu of the canonic TT, we have obtained an explicit expression for the GW strain in terms of amplitudes in the polarization matrix explicitly carrying time-influenced (and influenc*ing*) components, while preserving the rank-2 tensorial nature of the GW.

We have proposed an approach to the detection problem based on the observation of GW-induced time dilation: we have presented the concept design for an ideal detector, in which a physical process of know duration is timed after the GW has left the entire system made of source and timer. While this idea is constructed as the sum of many ansätze, a real world experimental observatory will have to deal with the technological limitations in terms of a sufficiently constant rate of signal emission, the capability of preserving the inter-signal separation, as well as a sufficiently precise detecting clock, with the non-obvious question regarding the versatility of such a system to become part of a feasible GW detection system in the vicinity of Earth, where the background metric is not Minkowski. This is left for future work as part of a feasibility study. The present experimental data on GW detection provide the constraints on the required clock resolutions needed to observe the same GW sources with our proposed system, or in other words, the amplitude of the temporal strain is the same order of magnitude as the spatial one. Other than this, there are no further constraints on our proposal coming from the current GW observations, since both gauges are constructed inside the boundaries of GR. In the final part of the paper, we described a possible realistic astrophysical scenario, an extreme mass ratio collision, in which the conditions for the generation of GWs appropriate for their detection with time dilation are immediately met. The natural extension of this concept to the detection of periodic GW signals has been touched briefly, but requires further development, mostly on technical grounds, and is left for future research.

**Acknowledgements** S.B. thanks Rai Weiss for the helpful discussion, insightful advice, and comments, and for the precious encouragement on the early ideas behind this work. S.B. also thanks Fabio Rigamonti for his intuition and suggestions about the experimental design.

**Data Availability Statement** This manuscript has no associated data or the data will not be deposited. [Authors' comment: This manuscript is a theoretical model and has no associated data. All the figures are either qualitative or refer directly to equations and parameters as they appear in the text.]

## References

1. LIGO Scientific Collaboration, Advanced LIGO. Class. Quantum Gravity **32**(7), 074001 (2015). https://doi.org/10.1088/0264-9381/32/7/074001. arXiv:1411.4547 [gr-qc]
2. F. Acernese et al., Advanced Virgo: a second-generation interferometric gravitational wave detector. Class. Quantum Gravity **32**(2), 024001 (2015). https://doi.org/10.1088/0264-9381/32/2/024001. arXiv:1408.3978 [gr-qc]
3. B.P. Abbott et al., Observation of gravitational waves from a binary black hole merger. Phys. Rev. Lett. **116**, 061102 (2016). https://doi.org/10.1103/PhysRevLett.116.061102
4. B.P. Abbott et al., Gwtc-1: a gravitational-wave transient catalog of compact binary mergers observed by ligo and virgo during the first and second observing runs. Phys. Rev. X **9**, 031040 (2019). https://doi.org/10.1103/PhysRevX.9.031040
5. R. Abbott et al., Gwtc-2: compact binary coalescences observed by ligo and virgo during the first half of the third observing run. Phys. Rev. X **11**, 021053 (2021). https://doi.org/10.1103/PhysRevX.11.021053
6. R. Abbott et al., Observation of gravitational waves from two neutron star-black hole coalescences. Astrophys. J. Lett. (2021). https://doi.org/10.3847/2041-8213/ac082e
7. A. Ashtekar, B. Bonga, On the ambiguity in the notion of transverse traceless modes of gravitational waves. Gen. Relativ. Gravit. (2017). https://doi.org/10.1007/s10714-017-2290-z
8. A. Ashtekar, B. Bonga, On a basic conceptual confusion in gravitational radiation theory. Class. Quantum Gravity **34**(20), 20–01 (2017). https://doi.org/10.1088/1361-6382/aa88e2
9. F. Canfora, L. Parisi, G. Vilasi, Nonlinear gravitational waves, their polarization, and realistic sources. Theor. Math. Phys. **152**(2), 1069–1080 (2007). https://doi.org/10.1007/s11232-007-0091-3
10. F. Canfora, G. Vilasi, Spin-1 gravitational waves and their natural sources. Phys. Lett. B **585**(3), 193–199 (2004). https://doi.org/10.1016/j.physletb.2004.02.005
11. B.P. Abbott et al., First search for nontensorial gravitational waves from known pulsars. Phys. Rev. Lett. **120**, 031104 (2018). https://doi.org/10.1103/PhysRevLett.120.031104
12. P. Amaro-Seoane et al., Laser interferometer space antenna (2017). arXiv:1702.00786 [astro-ph.IM]
13. W. Folkner, R. Hellings, L. Maleki, P. Bender, J. Fallen, R. Stebbins, K. Danzmann, J. Cornelisse, Y. Jafry, R. Reinhard, LISA-Laser Interferometer Space Antenna for gravitational wave measurements, in *33rd Aerospace Sciences Meeting and Exhibit* (1996). https://doi.org/10.2514/6.1995-829
14. E.S. Research, T. Centre, LISA Science Requirements Document. (ESA-L3-EST-SCI-RS-001) (2018). https://www.cosmos.esa.int/documents/678316/1700384/SciRD.pdf/25831f6b-3c01-e215-5916-4ac6e4b306fb?t=1526479841000
15. A. Sesana et al., Unveiling the gravitational universe at micro-hz frequencies (2019). arXiv:1908.11391 [astro-ph.IM]
16. B.P. Abbott et al. and KAGRA Collaboration, LIGO Scientific Collaboration and Virgo Collaboration. Prospects for observing and localizing gravitational-wave transients with Advanced LIGO, Advanced Virgo and KAGRA. Living Rev. Relativ. 21(1), 3. (2018). https://doi.org/10.1007/s41114-018-0012-9
17. T. Akutsu et al., Overview of KAGRA: Detector design and construction history. Prog. Theor. Exp. Phys. 2021(5), 05A101, https://doi.org/10.1093/ptep/ptaa125
18. A. Buikema, C. Cahillane, G.L. Mansell, C.D. Blair, R. Abbott et al., Sensitivity and performance of the Advanced LIGO detectors in the third observing run. Phys. Rev. D **102**(6), 062003 (2020). https://doi.org/10.1103/PhysRevD.102.062003. arXiv:2008.01301 [astro-ph.IM]
19. B.P. Abbott, et al. L.S.C.Kagra Collaboration, VIRGO Collaboration, Prospects for observing and localizing gravitational-wave transients with Advanced LIGO, Advanced Virgo and KAGRA. Living Rev. Relativ. **21**(1), 3 (2018). https://doi.org/10.1007/s41114-018-0012-9. arXiv:1304.0670 [gr-qc]
20. D. Christodoulou, Nonlinear nature of gravitation and gravitational-wave experiments. Phys. Rev. Lett. **67**, 1486–1489 (1991). https://doi.org/10.1103/PhysRevLett.67.1486
21. J.W. Maluf, J.F. Rocha-Neto, S.C. Ulhoa, F.L. Carneiro, Plane gravitational waves, the kinetic energy of free particles and the memory effect. Gravit. Cosmol. **24**(3), 261–266 (2018). https://doi.org/10.1134/S020228931803009X. arXiv:1707.06874 [gr-qc]
22. V.B. Braginsky, K.S. Thorne, Gravitational-wave bursts with memory and experimental prospects. Nature **327**(6118), 123–125 (1987). https://doi.org/10.1038/327123a0
23. L. Blanchet, T. Damour, Hereditary effects in gravitational radiation. Phys. Rev. D **46**, 4304–4319 (1992). https://doi.org/10.1103/PhysRevD.46.4304
24. O. Boersma, D. Nichols, P. Schmidt, Forecasts for detecting the gravitational wave memory effect with advanced LIGO and Virgo. Phys. Rev. D Part. Fields Gravit. Cosmol. (2020). https://doi.org/10.1103/PhysRevD.101.083026
25. A. Sesana, Pulsar timing arrays and the challenge of massive black hole binary astrophysics, in *Gravitational Wave Astrophysics*. Astrophysics and Space Science Proceedings, vol. 40, p. 147 (2015). https://doi.org/10.1007/978-3-319-10488-1_13
26. A. Sesana, A. Vecchio, Gravitational waves and pulsar timing: stochastic background, individual sources and parameter estimation. Class. Quantum Gravity **27**(8), 084016 (2010). https://doi.org/10.1088/0264-9381/27/8/084016. arXiv:1001.3161 [astro-ph.CO]
27. J.P.W. Verbiest et al., The international pulsar timing array: first data release. MNRAS **458**(2), 1267–1288 (2016). https://doi.org/10.1093/mnras/stw347. arXiv:1602.03640 [astro-ph.IM]
28. J. Antoniadis et al., The International Pulsar Timing Array second data release: search for an isotropic gravitational wave background. Mon. Not. R. Astron. Soc. **510**(4), 4873–4887 (2022). https://doi.org/10.1093/mnras/stab3418. https://academic.oup.com/mnras/article-pdf/510/4/4873/42242297/stab3418.pdf
29. J.P.W. Verbiest, S. Osłowski, S. Burke-Spolaor, Pulsar timing array experiments, in *Handbook of Gravitational Wave Astronomy*, p. 4. Springer (2021). https://doi.org/10.1007/978-981-15-4702-7_4-1
30. R.N. Manchester, Pulsar timing arrays and their applications, in *Radio Pulsars: An Astrophysical Key to Unlock the Secrets of the Universe*, ed. by M. Burgay, N. D'Amico, P. Esposito, A. Pellizzoni, A. Possenti. American Institute of Physics Conference Series, vol. 1357, pp. 65–72 (2011). https://doi.org/10.1063/1.3615080
31. F.B. Estabrook, H.D. Wahlquist, Response of Doppler spacecraft tracking to gravitational radiation. Gen. Relativ. Gravit. **6**(5), 439–447 (1975). https://doi.org/10.1007/BF00762449
32. C. Overstreet, P. Asenbaum, J. Curti, M. Kim, M.A. Kasevich, Observation of a gravitational Aharonov–Bohm effect. Science **375**(6577), 226–229 (2022). https://doi.org/10.1126/science.abl7152
33. M.A Hohensee, B. Estey, P. Hamilton, A. Zeilinger, H. Müller, Force–Free gravitational redshift: proposed gravitational Aharonov-Bohm experiment. Phys. Rev. Lett. 108(23), 230404 (2012). https://doi.org/10.1103/PhysRevLett.108.230404
34. J. Audretsch, C. Lammerzahl, Neutron interference: general theory of the influence of gravity, inertia and space-time torsion. J. Phys. A: Math. Gen. 16, 2457 (1983). https://doi.org/10.1088/0305-4470/16/11/017
35. Y.-J. Wang, X.-Y. Lu, C.-G. Qin, Y.-J. Tan, C.-G. Shao, Modeling gravitational wave detection with atom interferometry. Class. Quantum Gravity **38**(14), 145025 (2021). https://doi.org/10.1088/1361-6382/ac0236

36. A. Roura, Quantum probe of space-time curvature. Science **375**(6577), 142–143 (2022). https://doi.org/10.1126/science.abm6854
37. A. Shapere, F. Wilczek, Classical time crystals. Phys. Rev. Lett. **109**, 160402 (2012). https://doi.org/10.1103/PhysRevLett.109.160402
38. T. Li, Z.-X. Gong, Z.-Q. Yin, H.T. Quan, X. Yin, P. Zhang, L.-M. Duan, X. Zhang, Space-time crystals of trapped ions. Phys. Rev. Lett. **109**, 163001 (2012). https://doi.org/10.1103/PhysRevLett.109.163001
39. N.Y. Yao, A.C. Potter, I.-D. Potirniche, A. Vishwanath, Discrete time crystals: rigidity, criticality, and realizations. Phys. Rev. Lett. **118**, 030401 (2017). https://doi.org/10.1103/PhysRevLett.118.030401
40. A. Pizzi, A. Nunnenkamp, J. Knolle, Classical prethermal phases of matter. Phys. Rev. Lett. **127**, 140602 (2021). https://doi.org/10.1103/PhysRevLett.127.140602
41. A. Pizzi, A. Nunnenkamp, J. Knolle, Classical approaches to prethermal discrete time crystals in one, two, and three dimensions. Phys. Rev. B **104**, 094308 (2021). https://doi.org/10.1103/PhysRevB.104.094308
42. A.D. Ludlow, M.M. Boyd, J. Ye, E. Peik, P.O. Schmidt, Optical atomic clocks. Rev. Mod. Phys. **87**(2), 637–701 (2015). https://doi.org/10.1103/RevModPhys.87.637. arXiv:1407.3493 [physics.atom-ph]
43. M. Hena Mustary, L. Xu, W. Wu, N. Haram, D.E. Laban, H. Xu, F. He, I.V. Litvinyuk, R.T. Sang, Attosecond delays of high harmonic emissions from isotopes of molecular hydrogen measured by Gouy phase XUV interferometer (2021). arXiv:2111.05497 [physics.atom-ph]
44. D.E. Laban, A.J. Palmer, W.C. Wallace, N.S. Gaffney, R.P.M.J.W. Notermans, T.T.J. Clevis, M.G. Pullen, D. Jiang, H.M. Quiney, I.V. Litvinyuk, D. Kielpinski, R.T. Sang, Extreme ultraviolet interferometer using high-order harmonic generation from successive sources. Phys. Rev. Lett. **109**, 263902 (2012). https://doi.org/10.1103/PhysRevLett.109.263902
45. S.M. Brewer, J.-S. Chen, A.M. Hankin, E.R. Clements, C.W. Chou, D.J. Wineland, D.B. Hume, D.R. Leibrandt, $^{27}Al^+$ quantum-logic clock with a systematic uncertainty below $10^{-18}$. Phys. Rev. Lett. **123**, 033201 (2019). https://doi.org/10.1103/PhysRevLett.123.033201
46. G.E. Marti, R.B. Hutson, A. Goban, S.L. Campbell, N. Poli, J. Ye, Imaging optical frequencies with $100\mu$ Hz precision and $1.1\mu$ m resolution. Phys. Rev. Lett. **120**, 103201 (2018). https://doi.org/10.1103/PhysRevLett.120.103201
47. L. Bocklage, J. Gollwitzer, C. Strohm, C.F. Adolff, K. Schlage, I. Sergeev, O. Leupold, H.-C. Wille, G. Meier, R. Röhlsberger, Coherent control of collective nuclear quantum states via transient magnons. Sci. Adv. **7**(5), 3991 (2021). https://doi.org/10.1126/sciadv.abc3991
48. T. Bothwell, C.J. Kennedy, A. Aeppli, D. Kedar, J.M. Robinson, E. Oelker, A. Staron, J. Ye, Resolving the gravitational redshift across a millimetre-scale atomic sample. Nature **602**(7897), 420–424 (2022). https://doi.org/10.1038/s41586-021-04349-7'. arXiv:2109.12238 [physics.atom-ph]
49. V.O. Rivelles, Solitonic gravitational waves. Rev. Bras. Fis. **13**, 369–373 (1983)
50. B.J. Carr, E. Verdaguer, Soliton solutions and cosmological gravitational waves. Phys. Rev. D **28**, 2995–3006 (1983). https://doi.org/10.1103/PhysRevD.28.2995
51. B.C. Nichol, R. Srinivas, D.P. Nadlinger, P. Drmota, D. Main, G. Araneda, C.J. Ballance, D.M. Lucas, An elementary quantum network of entangled optical atomic clocks. Nature **609**(7928), 689–694 (2022). https://doi.org/10.1038/s41586-022-05088-z. arXiv:2111.10336 [physics.atom-ph]
52. GRAVITY Collaboration, A geometric distance measurement to the Galactic center black hole with 0.3% uncertainty. Astron. Astrophys. **625**, 10 (2019). https://doi.org/10.1051/0004-6361/201935656. arXiv:1904.05721 [astro-ph.GA]
53. Y.B. Zel'dovich, I.D. Novikov, The hypothesis of cores retarded during expansion and the hot cosmological model. Sov. Astron. **10**, 602 (1967)
54. Ya.B. Zeldovich, I.D.N., Probability for primordial black holes. Astronomicheskij Zhurnal **43** (1966)
55. S. Hawking, Gravitationally collapsed objects of very low mass. MNRAS **152**, 75 (1971). https://doi.org/10.1093/mnras/152.1.75
56. B.J. Carr, S.W. Hawking, Black holes in the early universe. Mon. Not. R. Astron. Soc. **168**(2), 399–415 (1974). https://doi.org/10.1093/mnras/168.2.399. https://arxiv.org/abs/https://academic.oup.com/mnras/article-pdf/168/2/399/8079885/mnras168-0399.pdf
57. R. Bousso, S.W. Hawking, Probability for primordial black holes. Phys. Rev. D **52**, 5659–5664 (1995). https://doi.org/10.1103/PhysRevD.52.5659
58. B. Carr, K. Kohri, Y. Sendouda, J. Yokoyama, Constraints on primordial black holes (2021)
59. B. Carr, S. Clesse, J. García-Bellido, F. Kühnel, Cosmic conundra explained by thermal history and primordial black holes. Phys. Dark Univ. **31**, 100755 (2021). https://doi.org/10.1016/j.dark.2020.100755
60. M. Maggiore, Gravitational Waves: Volume 1: Theory and Experiments (OUP Oxford, 2007) **(ISBN 0198570740)**
61. S. Weinberg, Gravitation and Cosmology: Principles and Applications of the General Theory of Relativity, 1st edn. (Wiley, 1972) **(ISBN 0-471-92567-5)**